# Communicating Using an Energy Harvesting Transmitter: Optimum Policies Under Energy Storage Losses

Kaya Tutuncuoglu Aylin Yener
Wireless Communications and Networking Laboratory
Electrical Engineering Department
The Pennsylvania State University, University Park, PA 16802
*kaya@psu.edu* *yener@ee.psu.edu*

***Abstract*—In this paper, short-term throughput optimal power allocation policies are derived for an energy harvesting transmitter with energy storage losses. In particular, the energy harvesting transmitter is equipped with a battery that loses a fraction of its stored energy. Both single user, i.e. one transmitter-one receiver, and the broadcast channel, i.e., one transmitter-multiple receiver settings are considered, initially with an infinite capacity battery. It is shown that the optimal policies for these models are threshold policies. Specifically, storing energy when harvested power is above an upper threshold, retrieving energy when harvested power is below a lower threshold, and transmitting with the harvested energy in between is shown to maximize the weighted sum-rate. It is observed that the two thresholds are related through the storage efficiency of the battery, and are non-decreasing during the transmission. The results are then extended to the case with finite battery capacity, where it is shown that a similar double-threshold structure arises but the thresholds are no longer monotonic. A dynamic program that yields an optimal online power allocation is derived, and is shown to have a similar double-threshold structure. A simpler online policy is proposed and observed to perform close to the optimal policy.**

## I. Introduction

Desirable aspects of future wireless applications include longer lifetime, smaller physical size, energy independence and a low carbon footprint. Energy harvesting wireless networks play an important role towards providing these, allowing mobile devices to operate for an indefinite amount of time. On the other hand, networks comprising of energy harvesting nodes have their own design challenges, most prominently, efficient use of intermittent energy taking into consideration the availability and storage of harvested energy.

Energy harvesting wireless networks have begun to receive attention from the wireless communication community. Efforts in the past few years have considered such networks and identified optimal policies to govern the scarce and varying energy resource. The transmission completion time minimization problem for an energy harvesting transmitter is considered in [2] with discrete energy arrivals known in an offline manner, and the policy satisfying the energy constraints is shown to have a non-decreasing piecewise constant structure. In [3], an energy harvesting node with limited energy storage is studied for throughput maximization, and a similar piecewise constant policy is found to be optimal. It is also shown in reference [3] that the problems of time minimization and throughput maximization are closely related. This study is extended to a single-link fading channel in [4], where a directional water-filling algorithm with varying water levels is proposed to combine the energy constraints with the conventional water-filling results for fading channels. A similar water-filling result is obtained by [5] with an information theoretic approach. Multiple user settings for offline problems have been studied in [6] for broadcast channels, [7] for multiple access channels and [8] for interference channels using variations of the directional water-filling algorithm. A two-hop setting with energy harvesting transmitter and relay in a static channel has been considered in [9]. All of these aforementioned efforts assume the presence of an energy storage unit on the node which is able to store the harvested energy without loss.

An energy storage device proves to be useful in designing more flexible power policies by providing a buffer for the harvested energy. Essentially, this helps prolong the operation of the node since the stored energy can be used whenever the node is needed on the network. However, the said storage device in reality would have non-ideal characteristics, such as capacity[1] fading, energy-expenditure-rate dependent capacity, leakage, and recovery effects. As a consequence, it is necessary to consider these imperfections to develop a more realistic model for wireless nodes and find power policies tailored to them. Various models have been proposed to predict the behavior of energy storage devices such as chemical batteries [10], [11]. Energy harvesting nodes utilizing batteries with capacity fading or battery leakage were studied in [12] by revising the approaches of [2], [3]. Storage inefficiency was modeled in [13] as a constant loss rate per stored energy to find asymptotically optimal policies for sufficiently large batteries with energy neutrality constraints. Reference [14] studied duty-cycling with constant transmission rate under energy neutrality conditions.

This paper focuses on single transmitter communication settings where the transmitter is energy harvesting, and a fraction of the (harvested) energy is wasted due to imperfections in storing in as well as discharging energy from the energy storage device on board, such as a battery[2] or supercapacitor. First, optimal offline policies maximizing the average rate are found for a single transmitter-receiver link, and are also shown to solve the broadcast channel, with a sufficiently large battery capacity, in Sections III and IV respectively. It is shown that contrary to the results of previous work with ideal batteries [2]–[4], [6], [7], where the optimal policy is shown to be piece-wise constant, here, the optimal policy for a battery with storage losses may favor transmitting with the just harvested energy without storing it. In particular, the optimal policy is shown to have a double-threshold structure with non-decreasing thresholds, in which the transmit power equals the harvested power whenever the harvested power falls between the thresholds. Next, the results are extended to the case when the storage device is a finite capacity battery in Section V. For this case, it is shown that the double threshold policy applies, while the thresholds are instead determined as piecewise constant and decreasing or increasing at full and empty battery instances, resembling the power policy in [3]. Building on the intuition from the optimal offline policy, a dynamic program to find the optimal online policy is presented and a simpler near-optimal policy with constant threshold levels maintaining a stable energy buffer is proposed in Section VI. The performances of these policies are simulated and compared to the offline policy in Section VII.

[1]Here we refer to *energy storage capacity* of said device. The data transmission related metrics are referred to as short-term throughput or rate.

## II. System Model

We consider an energy harvesting transmitter that employs transmission power control to regulate the achieved rate or utility. The node is free to choose how much of the harvested energy will be utilized for transmission, storing the remaining portion in the on-board battery. The instantaneous transmission power is drawn from the energy being harvested at that instant, the energy previously stored in the battery, or both, depending on the transmission policy.

The instantaneous utility $r(p)$ is defined as the utility when the node transmits with power $p$. The utility of the system is then defined as the integral of the achieved instantaneous utility over the duration of operation, $T$. For the single link, $r(p)$ can be the achieved instantaneous rate, with the system utility translating to the total number of bits communicated to the receiver. For the broadcast setting, $r(p)$ can be any weighted sum of the number of bits delivered to all the receivers.

The node harvests energy at a non-negative rate $h(t)$ to either be used in transmission directly or to be stored in the battery. However, due to the inefficiency of the battery, a fraction of the stored power is lost. This energy loss model has been used before, see for example, [13]. Similarly, a loss may occur when power is drawn from the battery. These two losses are combined in the model and the fraction of energy that can be drawn from the battery per unit energy stored, i.e., battery efficiency, is represented by $\eta$, $0 \leq \eta \leq 1$. The model for the single user setting is depicted in Figure 1.

[2]From this point on we use battery and energy storage device interchangeably.

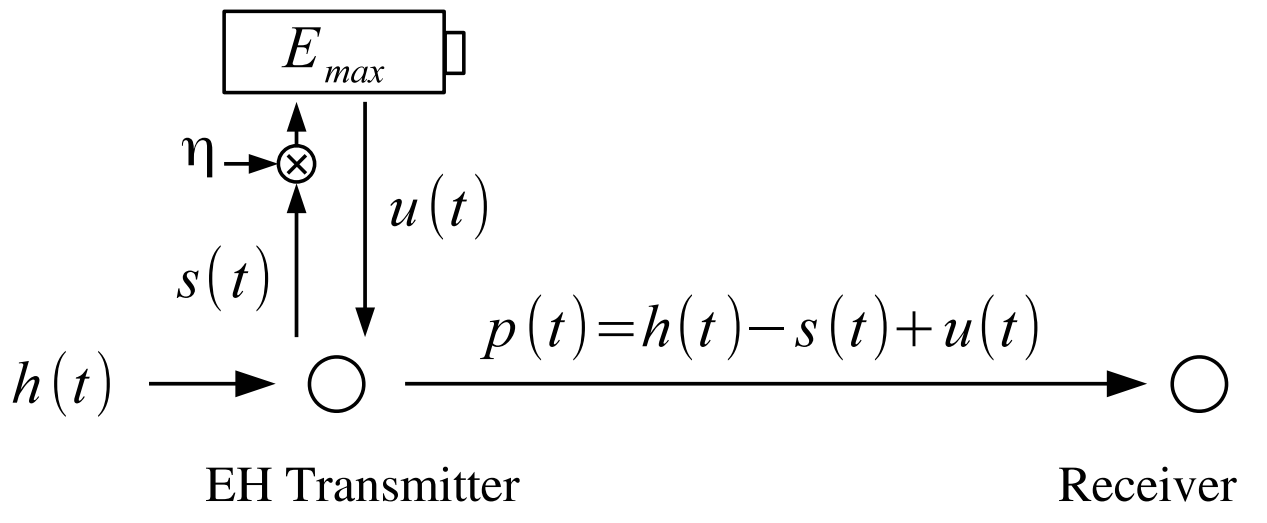

Fig. 1. Energy harvesting transmitter with inefficient storage and finite battery capacity in a single link.

Denoting the rate of energy storage by $s(t)$ and the power drawn from the battery by $u(t)$, the energy stored in the battery at time $t$ is given by

$$E_{bat}(t) = \int_0^t \eta s(\tau) - u(\tau)\ d\tau \tag{1}$$

and the power the transmitter receives, which is the remaining harvested power and the power drawn from the battery, is expressed as

$$p(t) = h(t) - s(\tau) + u(\tau). \tag{2}$$

Note that the power being stored cannot exceed the harvested power, i.e., $s(t) \leq h(t)$, and both $s(t)$ and $u(t)$ are non-negative by definition.

It is worthwhile to note that $s(t)$ and $u(t)$ should not be nonzero at the same time, since storing and drawing energy simultaneously may not be physically possible [15], and also that it yields to an energy loss without any storage benefit. We will not impose such a constraint on the problem, but it will become evident that the optimal policy satisfies this constraint. Secondly, the energy drawn from the battery cannot exceed the energy stored in the battery at any time throughout the transmission. This constraint, which we will refer to as *energy causality*, is given for a total transmission duration of $T$ by the set of constraints

$$E_{bat}(t) = \int_0^t \eta s(\tau) - u(\tau)\ d\tau \geq 0, \qquad 0 \leq t \leq T. \tag{3}$$

In Section V, we shall consider nodes with a finite storage capacity of $E_{max}$. A finite capacity implies that any power attempted to be stored in the battery when $E_{bat} = E_{max}$ will be lost. This loss on the battery side can be avoided by directing the excess power to the transmitter, yielding *battery capacity* constraints

$$E_{bat}(t) = \int_0^t \eta s(\tau) - u(\tau)\ d\tau \leq E_{max}, \qquad 0 \leq t \leq T \tag{4}$$

that prevent battery overflow. Note that this constraint does not allow the storage system to discard any energy. However, in the unlikely case that it is optimal to do so, the discard decision is left to the transmitter through the design of the utility function $r(p)$.

The problem investigated in this paper is maximizing the average utility of the system within a deadline $T$ through a transmission power adapted to the harvesting process and the inefficient storage, for a single energy harvesting transmitter and single receiver (Section III), and multiple receiver (Section IV) settings. We shall solve this problem in an offline setting, i.e., the energy harvests known to the transmitter. Using the insights obtained from the optimal offline policies, we shall propose online policies in Section VI.

## III. Optimal Transmission Policy for a Single Link

The average utility maximization problem for the model in Figure 1 with infinite battery capacity is defined as

$$\max_{s(t),u(t)} \frac{1}{T}\int_0^T r(h(t)-s(t)+u(t))\ dt \tag{5a}$$

$$\text{s.t. } 0 \leq \int_0^t \eta s(\tau)-u(\tau)\ d\tau,$$
$$h(t)\geq s(t)\geq 0,\ u(t)\geq 0, \quad 0\leq t\leq T \tag{5b}$$

where $\eta$ is the efficiency of the battery for the energy harvesting transmitter. We first observe that the problem in (5) is a convex optimization problem. The constraints on $s(t)$ and $u(t)$ in (5b) are linear, and thus the feasible set is convex. To show the convexity of the problem as a whole, it remains to show that the objective function in (5a) is concave in these variables. First, we state the concavity of $r(p)$ in $p$ through a time-sharing argument similar to that in [8]:

*Lemma 1:* The maximum achievable instantaneous rate $r(p)$ for a given power $p$ is non-decreasing, continuous and concave in $p$.

*Proof:* The non-decreasing property emerges from the ability of the transmitter to discard a fraction of the provided transmission power $p$ and achieve the rate provided by any $p' \leq p$. Continuity follows from the following observation: given a power $p' < p$ arbitrarily close to $p$, the transmitter can achieve a rate arbitrarily close to $r(p)$ by transmitting with $r(p)$ for a slightly shorter time period and turning off for a sufficient time so that an average transmit power of $p'$ is achieved.

The proof for concavity is by contradiction. For any $p_1$, $p_2$ and $\lambda$, assume the concavity is violated at $p_\lambda = \lambda p_1 + (1-\lambda)p_2$. Then it is easy to see that a rate better than $r(p_\lambda)$ can be achieved by time-sharing between $r(p_1)$ and $r(p_2)$ with a sharing parameter $\lambda$. Consequently, $r(p)$ cannot be the maximum achievable instantaneous rate. ■

Essentially Lemma 1 suggests that the desired properties all follow from an efficient use of the available instantaneous power. If it is possible to achieve a better rate with less power, some energy can be discarded. If it is more efficient to allow the node to sleep and wake-up, this can be considered within the instantaneous rate function by performing the corresponding sleep policy when needed. Similarly, if time-sharing between two or more power levels with an average power of $p$ achieves a better rate, the node would adopt this time-sharing policy whenever supplied with a power of $p$. Benefits these simple policies might bring are included in the power-rate function, which also renders this function non-decreasing, continuous and concave.

*Corollary 1:* Since $p(t) = h(t)-s(t)+u(t)$ is a linear function of $s(t)$ and $u(t)$, the objective function in (5a) is continuous and jointly concave in the variables of the problem.

The concavity of the objective function of the maximization and the convexity of the constraint set implies that (5) is a convex program. The Lagrangian corresponding to (5) is given in (6), where $\lambda(t)$, $\mu(t)$, $\sigma(t)$ and $\nu(t)$ are the nonnegative Lagrangian multipliers corresponding to the energy causality, $s(t)\geq h(t)$, and nonnegativity constraints for $s(t)$ and $u(t)$ respectively. The optimal energy storage and use policy must satisfy the Karush-Kuhn-Tucker (KKT) stationarity conditions [16] found by taking the derivative with respect to both variables at time $0\leq t\leq T$ as

$$r'(h(t)-s(t)+u(t)) - \eta\int_t^T \lambda(\tau)\ d\tau + \mu(t) - \sigma(t) = 0 \tag{7}$$

$$-r'(h(t)-s(t)+u(t)) + \int_t^T \lambda(\tau)\ d\tau - \nu(t) = 0 \tag{8}$$

for $0\leq t\leq T$ where $r'(p)$ represents the derivative of $r(p)$ with respect to $p$. The corresponding complementary slackness conditions for each Lagrangian multiplier are

$$\lambda(t)\left(\int_0^t \eta s(\tau)-u(\tau)\ d\tau\right) = 0 \tag{9a}$$

$$\mu(t)\left(h(t)-s(t)\right) = 0 \tag{9b}$$

$$\sigma(t)s(t) = 0, \qquad \nu(t)u(t) = 0 \qquad 0\leq t\leq T. \tag{9c}$$

In order to find the optimal policy, we test the KKT conditions above for five mutually exclusive modes of the transmitter that include all possible choices of $s(t)$ and $u(t)$. In cases where $r(p)$ is not strictly concave, such as when time-sharing is employed, the solution of this problem is not unique. To develop an algorithm with a unique output and to simplify the analysis, we restrict our search set by omitting the modes which are strictly suboptimal or can be replaced with another mode without loss of optimality.

*Case 1: Simultaneous charge and discharge*

In this case, the battery is being charged and discharged at the same time $t$, i.e., $s(t) > 0$ and $u(t) > 0$. Due to the complementary slackness conditions in (9c), this implies that both $\sigma(t) = 0$ and $\nu(t) = 0$. Substituting these in (7) and (8) and adding the two equations, we get

$$(1-\eta)\int_t^T \lambda(\tau)\ d\tau + \mu(t) = 0. \tag{10}$$

Due to $\mu(t)$ and $\lambda(t)$ being non-negative, and $0\leq\eta\leq 1$, the conditions for (10) to hold are

$$\mu(t) = 0 \tag{11}$$

$$(1-\eta)\int_t^T \lambda(\tau)\ d\tau = 0. \tag{12}$$

(12) implies that for the transmitter to be in this mode, either the efficiency $\eta$ needs to be 1, or $\lambda(\tau) = 0$ for all $t\leq\tau\leq T$. In the former case, the storage is lossless, and a simultaneous charge and discharge is equivalent to only charging or only discharging with $\min(s(t),u(t))$ forwarded directly from harvested power to the transmitter. The latter case, substituted in (8) gives $r'(p(\tau)) = 0$ for all $t\leq\tau\leq T$ with $u(t) = 0$, meaning that the rate is invariant to transmission power after $t$ whenever the stored energy is used. Thus, the expended power

$$\mathcal{L} = \int_0^T r(h(t) - s(t) + u(t)) \ dt + \int_0^T \left( \lambda(t) \int_0^t \eta s(\tau) - u(\tau) \ d\tau + \mu(t)(h(t) - s(t)) + \sigma(t)s(t) + \nu(t)u(t) \right) dt \tag{6}$$

is useless. This energy can equivalently be lost by increasing transmission power without getting more rate. Therefore, for both of the cases when simultaneous charge and discharge appears optimal, there exists an equally good policy that avoids this mode. Consequently, we can safely assume that this mode is never used by the transmitter.

*Case 2: Discharging only*

Since the simultaneous charge and discharge is considered in Case 1, this case strictly refers to $u(t) > 0$ and $s(t) = 0$. In this case, the complementary slackness condition in (9c) gives $\nu(t) = 0$, and substituting in (8) yields

$$r'(p(t)) = \int_t^T \lambda(\tau) \ d\tau. \tag{13}$$

Note that due to (9a), $\lambda(t)$ is nonzero only when the battery is empty. Hence, the value of $r'(p(t))$ in the discharging mode remains constant unless the battery is depleted. Since there are no other restrictions on $p(t)$ for optimality, one solution is to choose a constant transmission power $p = p_u$ as the smallest power satisfying (13), which we shall adopt for simplicity and ease of implementation in this paper. The reason behind choosing the smallest such value will become clear in Case 5.

*Case 3: Charging only with $s(t) < h(t)$*

In Case 3 and 4, we consider the charging mode in two parts with $s(t) < h(t)$ and $s(t) = h(t)$ respectively. For the case with $0 < s(t) < h(t)$ and $u(t) = 0$, i.e., the harvest rate is strictly larger than the storage rate and thus transmission power is nonzero, complementary slackness conditions (9b) and (9c) dictate that $\sigma(t) = 0$ and $\mu(t) = 0$. Substituting in (7), we get

$$r'(p(t)) = \eta \int_t^T \lambda(\tau) \ d\tau. \tag{14}$$

Noticing the similarity of the above equation to (13), we observe that $r'(p(t))$ in the charging mode also remains constant while the battery is not depleted, and introduce a similar restriction to the solution by choosing the largest constant transmission power $p = p_s$ satisfying (14) in this mode. We also note that the transmission power in discharge mode $p_u$ and the transmission power in the charge mode $p_s$ are related through

$$\frac{r'(p_s)}{r'(p_u)} = \eta \tag{15}$$

by (13) and (14). Therefore, we can conclude that there exists an optimal policy which charges and discharges only while maintaining constant transmission powers $p_s$ and $p_u$, and that these two powers are related by (15).

*Case 4: Charging only with $s(t) = h(t)$*

In this case, we consider storing all harvested power, i.e., $s(t) = h(t)$, and thus having no transmit power. Since the constraint $h(t) \geq s(t)$ is met with equality in this case, the corresponding Lagrangian multiplier $\mu(t)$ no longer has to be zero as in Case 3. Thus, (7) becomes

$$r'(p(t)) = r'(0) = \eta \int_t^T \lambda(\tau) \ d\tau - \mu(t) \tag{16}$$

where $p(t) = 0$ for this case by definition. Comparing to (14), $r'(0)$ is not greater than $r'(p_s)$ since $\mu(t) \geq 0$. Lemma 1 shows that $r(p)$ is non-decreasing continuous and concave, implying that $0 \geq p_s$ which is only feasible when $p_s = 0$. Therefore, this mode is only optimal when the transmission power for charging, defined as $p_s$ in the previous case, is zero. As a result, these two modes can be considered jointly as the charging mode, with the transmission power equal to $p_s$ found from (14).

*Case 5: No charging or discharging*

This is the case where the node forwards all harvester power to the transmitter, i.e., $p(t) = h(t)$ with $s(t) = u(t) = 0$, which we shall refer to as the *passive storage* mode. We assume that $h(t) > 0$ to avoid the trivial case of $h(t) = s(t) = u(t) = p(t) = 0$. In this case, substituting $\mu(t) = 0$ in (7) and (8) gives

$$r'(p(t)) = \eta \int_t^T \lambda(\tau) \ d\tau + \sigma(t) = r'(p_s) + \sigma(t) \leq r'(p_s) \tag{17a}$$

$$r'(p(t)) = \int_t^T \lambda(\tau) \ d\tau + \nu(t) = r'(p_u) - \nu(t) \geq r'(p_u) \tag{17b}$$

implying that the transmission power $p(t)$ is restricted to be within the interval $[p_u, p_s]$. Notice that this is in part due to the selection of $p_u$ and $p_s$ as the smallest and largest power values satisfying (13) and (14) respectively.

The analysis of Cases 1 to 5 imply that there exists an optimal policy with the following three modes:

1) Charging only mode with $p(t) = p_s$ such that (14) is satisfied,
2) Discharging only mode with $p(t) = p_u$ such that (13) is satisfied, and
3) Passive storage mode with $p_u \leq p(t) \leq p_s$.

It is straightforward to see that the above spells a double threshold policy on $h(t)$. When $h(t) > p_s$, transmission power is chosen as $p_s$ and the excess energy is stored in the battery, referring to the first mode. Conversely when $h(t) < p_u$, transmission power is kept at $p_u$ with the missing energy supplied from the battery, referring to the second mode. In between the two thresholds, i.e., when $p_u \leq h(t) \leq p_s$, the node transmits with the harvested power without utilizing the energy storage by any means, referring to the third mode. This policy gives a unique power allocation, satisfying all KKT conditions and the assumptions given in above cases, so that it performs at least as good as any other policy satisfying the necessary conditions. Thus the resulting policy is optimum.

At this point, what remains to obtain an optimal power allocation scheme is to determine the values of the thresholds, i.e., $p_u(t)$ and $p_s(t)$, throughout the transmission period $[0,T]$. Recall that these thresholds are defined as an integral of the Lagrangian variable $\lambda(t)$ as in (13) and (14) respectively. Remembering that $r'(p)$ is also non-increasing in $p$, it can be stated that these thresholds are non-decreasing in $t$ due to $\lambda(t) \geq 0$, and remain constant as long as the battery is non-empty due to (9a). Moreover, due to the relation in (15) and the definitions of $p_u$ and $p_s$, given a threshold $p_u(t)$ the corresponding $p_s(t)$ can be uniquely found and vice versa. Therefore, it suffices to determine a non-decreasing $p_u(t)$ that only changes at times of $E_{bat}(t)=0$ to find the optimal policy.

In order to find a threshold function $p_u(t)$ satisfying the properties above, we first make the following observation parallel to [3] and [2]:

*Lemma 2:* There exists an optimal transmission policy that terminates at $t=T$ with an empty battery, i.e., $E_{bat}(T)=0$.

The proof of this statement can be found in [3] and [2] as a necessary condition for optimality, and relates to $r(p)$ being strictly increasing. In our current set, this requirement is slightly relaxed to $r(p)$ being non-decreasing, thus, one might think that this termination requirement may not be necessary. However, Lemma 2 still holds since for any optimal transmission policy with $E_{bat}(T) > 0$, one can construct a policy performing at least as good by increasing the transmission power towards the end of transmission to deplete the battery at $T$.

We find the threshold function $p_u(t)$ by a one-dimensional search: Knowing that it is non-decreasing and constant while $E_{bat} > 0$, the smallest threshold $p_{u1} \geq 0$ and the corresponding $p_{s1} > p_{u1}$ that depletes the battery at some time $t_1 \leq T$ is found. These thresholds are set as $p_u(t)$ and $p_s(t)$ for $t \in [0,t_1]$. If $t_1 = T$, the algorithm terminates; otherwise, a next threshold $p_{u1} \geq p_{u2}$ that depletes the battery at a later time $t_1 < t_2 \leq T$ is found. This is repeated until the termination condition is met.

A sample optimal transmission policy is depicted in Figure 2. The harvesting process $h(t)$ represents a predictable solar harvesting pattern inspired from the example in [17, Figure 1]. The first set of thresholds $p_{u1}$ and $p_{s1}$ are determined as the smallest thresholds depleting the battery sometime in $[0,T]$, namely $t_1$ in this figure. The second set of thresholds $p_{u2}$ and $p_{s2}$ are determined starting from $t_1$ as the smallest ones depleting the battery in $[t_1,T]$, which consequently coincided with the deadline $T$. With these threshold values, the transmitter power $p(t)$ is shown in bold. Note that the lower threshold $p_{u2}$ is not effective until the battery is charged for the first time after $t_1$, since in this interval the battery is still empty, and the thresholds are free to change gradually while $E_{bat}=0$. The energy above $p(t)$, denoted with the shaded regions, are stored in the battery and used up to provide the energy denoted with the dotted regions.

*Remark 1:* The optimal policy derived in this section can be shown to converge to the results of [2] when the energy storage is assumed to be ideal. This is when $\eta = 1$, and any harvested power can be stored without a loss. The relationship

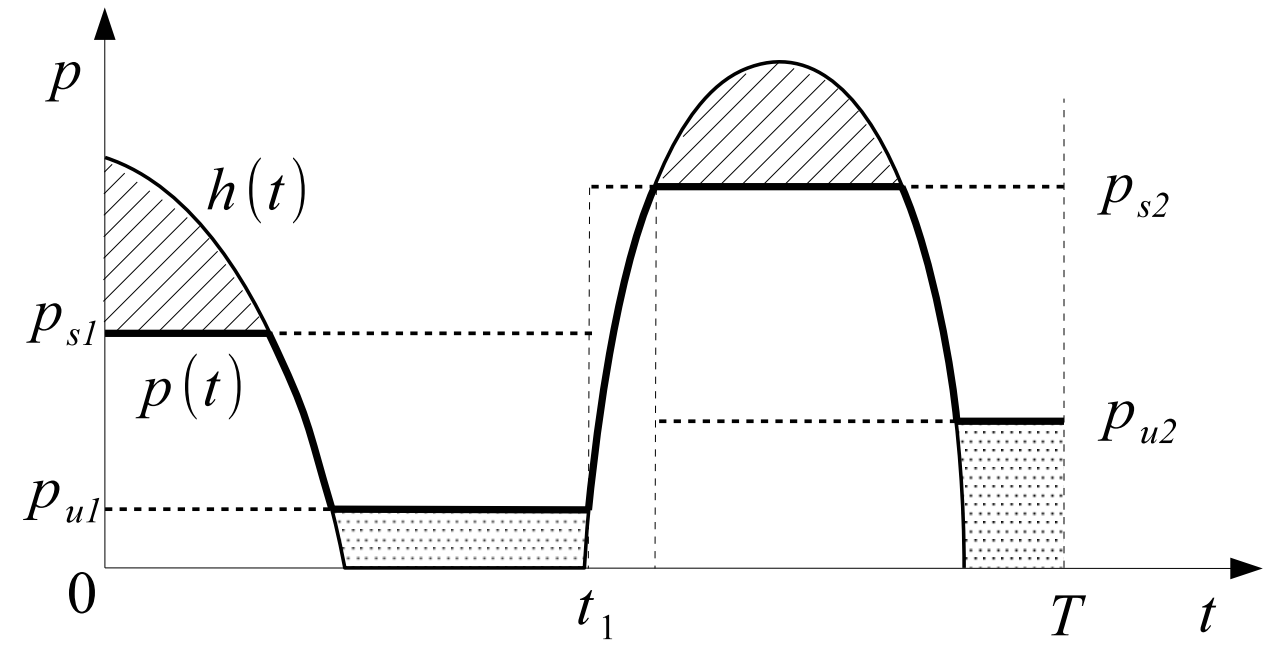

Fig. 2. Example optimal policy with transmission power thresholds $p_s$ and $p_r$ and a single empty battery instance at $t_1$

between the two thresholds, i.e., (15), becomes

$$\frac{r'(p_s)}{r'(p_u)} = \eta = 1 \tag{18}$$

When a strictly concave rate function is considered, such as in [2], the rate function is strictly increasing. Thus the only solution to (18) is at $p_s = p_u$, i.e., when the transmitter only transmits with constant power $p = p_s = p_u$, and stores or retrieves the harvested power accordingly. Consequently, the optimal policy consists of constant power transmissions, with the power level increasing only at instances of empty battery so that $\lambda(t) > 0$. Since the infinite battery transmission completion time minimization problem in [2] was shown to have an identical solution with the short-term throughput maximization problem we consider in reference [3], our solution coincides with that of [2] with ideal batteries.

## IV. Optimal Transmission Policy for the Broadcast Channel

In this section, we extend the results of Section III to the multi-receiver setting. For simplicity, we consider two receivers, although the results are easily generalizable to more than two. The channel model is depicted in Figure 3. For this setting, we wish to find an average rate region $\mathfrak{R}_{EH} = (r_{1,avg}, r_{2,avg})$ which is the union of average rate pairs that can be achieved under the energy harvesting constraints in (3) and (4).

At any time $t$, the transmitter allocates the power $p(t)$ for transmission, and can achieve any rate pair $(r_1,r_2) \in \mathfrak{R}(p(t))$ where $\mathfrak{R}(p(t))$ is the achievable rate region for transmit power $p(t)$. For example, for the static additive white Gaussian noise (AWGN) channel, the capacity region $\mathfrak{R}_{AWGN}(p)$, achieved by superposition coding, is known to be as in (19) when $\sigma_1 \leq \sigma_2$ [18]. For the AWGN channel, it is trivial that this region is convex for a fixed $p$. This property can be extended beyond this special case by pointing out the availability of time-sharing. If two points $(r_1,r_2) \in \mathfrak{R}$ and $(r_1',r_2') \in \mathfrak{R}$ are achievable, then by time-sharing the two schemes, any convex combination can also be achieved, and thus $\mathfrak{R}(p)$ is convex for a fixed $p$.

A more relevant property to the energy harvesting problem is the concavity of $\mathfrak{R}(p)$ in transmit power $p$. Specifically, if two rate pairs $(r_1,r_2) \in \mathfrak{R}(p)$ and $(r_1',r_2') \in \mathfrak{R}(p')$ can be

$$\mathfrak{R}_{AWGN}(p) = \left\{ (r_1, r_2) \middle| r_1 \leq \frac{1}{2}\log_2\left(1 + \frac{\alpha p}{\sigma_1^2}\right),\ r_2 \leq \frac{1}{2}\log_2\left(1 + \frac{(1-\alpha)p}{\alpha p + \sigma_2^2}\right),\ 0 \leq \alpha \leq 1 \right\} \tag{19}$$

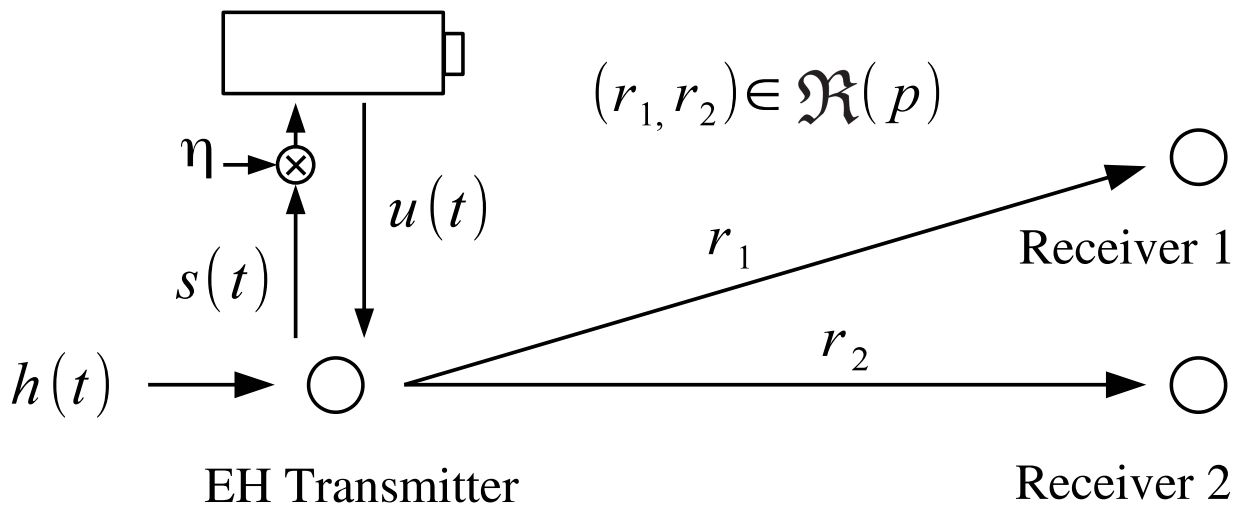

Fig. 3. Energy harvesting transmitter with inefficient storage in a broadcast setting.

achieved with transmit powers $p$ and $p'$ respectively, then their convex combination, i.e., $(\lambda r_1 + (1-\lambda)r_1', \lambda r_2 + (1-\lambda)r_2')$ must be within the achievable rate region for transmit power $\lambda p + (1-\lambda)p'$, where $0 \leq \lambda \leq 1$. Similar to the convexity of $\mathfrak{R}$, this argument follows from time-sharing between the two rate pairs and corresponding powers, and applies to any broadcast model in which time-sharing is applicable.

Combined with the linearity of the constraints on transmit power for an energy harvesting broadcaster, the observations above yield to the following Lemma:

*Lemma 3:* The achievable average rate region $\mathfrak{R}_{EH} = (r_{1,avg}, r_{2,avg})$ for an energy harvesting transmitter under power constraints (3) and (4) is convex.

*Proof:* It is to be shown that for any two achievable average rate pairs, all of their convex combinations are also achievable. Let $p(t)$ and $p'(t)$ be two feasible power allocation policies achieving rate pairs $(r_1(t), r_2(t))$ and $(r_1'(t), r_2'(t))$, yielding the average rate pairs $(r_{1,avg}, r_{2,avg})$ and $(r_{1,avg}', r_{2,avg}')$ respectively. A convex combination of these average rates with parameter $\lambda$ can be achieved by employing the transmit power $\lambda p(t) + (1-\lambda)p'(t)$ and choosing the rates as $(\lambda r_1(t) + (1-\lambda)r_1'(t), \lambda r_2(t) + (1-\lambda)r_2'(t))$. Note that these rates are achievable due to the concavity of $\mathfrak{R}(p)$ as discussed above. The feasibility of the power allocation follows from (2) and the constraints in (3) and (4) being linear, i.e., if two sets of $s(t)$ and $u(t)$ satisfy these constraints, then so does their convex combination. ∎

So far, we have shown that the achievable average rate region is convex. This is similar to the convexity of the maximum departure region in [6]. Consequently, one can trace its boundary by maximizing weighted sum rates, since each boundary point will be the maximizer to at least one set of weights. Moreover, by allocating all power to only one of the receivers, we can deduce that this region rests inside the box $[0, r_{1,max}] \times [0, r_{2,max}]$ where $r_{j,max}$ is the average rate achieved when all harvested power is used for transmitting to user $j$ as in Section III. Since three of the corner points of this box can trivially be achieved, we can further restrict our attention to boundary points maximizing the weighted sum $ar_1 + r_2$ with $a \geq 0$.

Given an instantaneous transmit power $p$, let the maximum achievable weighted sum-rate which satisfies $(r_1, r_2) \in \mathfrak{R}(p)$ be $r_a^{BC}(p)$. We define the average sum rate maximization problem, which allows to find the boundary of $\mathfrak{R}_{EH}$, as follows:

$$\max_{s(t),u(t)} \frac{1}{T}\int_0^T r_a^{BC}(h(t) - s(t) + u(t))\ dt \tag{20a}$$

$$\text{s.t. } 0 \leq \int_0^t \eta s(\tau) - u(\tau)\ d\tau,$$
$$h(t) \geq s(t) \geq 0, \quad u(t) \geq 0, \quad 0 \leq t \leq T. \tag{20b}$$

It turns out that the policy described in Section III extends to this setting in a straightforward manner. To show this, we begin by characterizing $r_a^{BC}(p)$.

*Lemma 4:* The maximum achievable weighted sum-rate $r_a^{BC}(p)$ for any coefficient $a \geq 0$ is non-decreasing, continuous and concave in $p$.

*Proof:* The proof is similar to that of Lemma 1. The non-decreasing property follows from the transmitter discarding excess energy to achieve any rate for a lower power. By time-sharing between an off state and any power $p$, one can get arbitrarily close to $r_a^{BC}(p)$ using power $p - \epsilon$ where $\epsilon > 0$ is arbitrarily small, showing continuity. Finally, time sharing between any pair of powers $p_1$ and $p_2$ with parameter $\lambda$ ensures that the rate function is concave within $[p_1, p_2]$. ∎

Comparing Lemma 1 and Lemma 4, we observe that the rate functions have the same properties which are sufficient to prove the optimality of the policy in Section III. The optimal broadcast channel power policy therefore has a double-threshold structure with the increasing thresholds found by a search, with the achieved weighted sum-rate at time $t$ given by $r_a^{BC}(p(t))$.

A fair question regarding this setting, and in fact for all multi-user models with more than one rate in the objective, is how to choose the individual rates for the users given the power of the broadcasting node. In a broadcast channel, the achievable rate tuples for a broadcast power is given by an achievable region. Due to the linear structure of the objective function, the optimal choice of the rate tuple arises as the one maximizing the weighted sum-rate for the given instantaneous power, found on the boundary of the achievable region having the weight ratio $a$ as a subgradient.

An interesting outcome is that the double threshold structure of the optimal policy is valid for any weight ratio $a$. That said, $a$ is a critical parameter of the system since the threshold values relate to $a$ through $r_a^{BC}(p)$ and (15); with the time-sharing coding scheme that achieves a particular point on $r_a^{BC}(p)$ also depending on the structure of the weighted sum-rate.

*Remark 2:* The results of this section can also be shown to converge to previous results on the broadcast channel in [6] when storage efficiency $\eta = 1$. In this case, the thresholds are once again found to be equal as in Remark 1, and total power levels constant and nondecreasing throughout the transmission are therefore found to be optimal, consistent with [6, Lemma 3] for an energy harvesting broadcast node with ideal and infinite energy storage.

## V. Extension to Finite Battery Models

In practice, it is likely that the storage device is of finite capacity, or it might be beneficial for design purposes to have as small a storage device as possible. Thus, it is relevant to consider the single user problem in (5) by including the battery capacity constraint in (4). The problem thus becomes

$$\max_{s(t),u(t)} \frac{1}{T}\int_0^T r(h(t)-s(t)+u(t))\ dt \tag{21a}$$

$$\text{s.t. } 0 \le \int_0^t \eta s(\tau)-u(\tau)\ d\tau \le E_{max},$$
$$h(t) \ge s(t) \ge 0, \quad u(t) \ge 0, \quad 0 \le t \le T \tag{21b}$$

With the added energy storage limitation of $E_{max}$, the Lagrangian of (21) becomes (22) where $\beta(t)$ is the non-negative Lagrangian multiplier for the new constraint, with the corresponding complementary slackness condition

$$\beta(t)\left(\int_0^t \eta s(\tau)-u(\tau)\ d\tau - E_{max}\right) = 0 \qquad 0 \le t \le T \tag{23}$$

in addition to the ones listed in (9). Substituting this modified Lagrangian in Cases 1 to 5 of Section III, we observe that the threshold values are still related with (15), and can be expressed as

$$r'(p_u(t)) = \int_t^T \lambda(\tau)\ d\tau - \int_t^T \beta(\tau)\ d\tau \tag{24}$$

$$r'(p_s(t)) = \eta\left(\int_t^T \lambda(\tau)\ d\tau - \int_t^T \beta(\tau)\ d\tau\right). \tag{25}$$

In the infinite battery case, non-negativity of $\lambda(t)$ implied non-decreasing threshold powers, and the complementary slackness condition in (9a) implied that the thresholds could only increase when the battery was empty. With the finite battery constraint, due to the added $\beta(t)$ terms in (24) and (25), this statement is revised as follows: The thresholds can only increase when the battery is empty, $E_{bat} = 0$, and can only decrease when the battery is full, $E_{bat} = E_{max}$. Similar to its counterpart, the second statement follows from the condition in (23), where either $\beta(t)$ or energy stored in the battery at time $t$ has to be zero at any given $t$.

What remains is to find an algorithm that gives the optimum threshold levels for $p_u$ or $p_s$ satisfying the above conditions. An optimal policy must follow the restrictions above while ensuring that the transmission terminates with $E_{bat}(T) = 0$ to avoid suboptimality due to energy loss. This statement provides a sufficient decision metric to find the optimal threshold levels. Consider that for some epoch $[t_0, t_1]$, the thresholds $p_s$ and $p_u$ is a candidate pair. For these thresholds to increase or decrease at $t_1$ and yield the next thresholds, the battery must be empty or full respectively. Assume first that $E_{bat}(t_1) = 0$, indicating that the thresholds will increase and thus less energy will be stored in the next epoch than what would have been stored if the same thresholds were to extend beyond $t_1$. Therefore, looking at the next battery event if $p_s$ and $p_u$ extended beyond $t_1$ gives important information about the possible threshold changes in the future. If this is another empty battery event, storing less energy with the next thresholds would yield an empty battery even earlier, and the node would be forced to transmit suboptimally. Conversely, assume that $E_{bat}(t_1) = E_{max}$, and next thresholds are thus less than the candidate thresholds, storing relatively more energy after $t_1$. If the next battery event for candidate thresholds is another full battery event, storing more energy would cause energy overflow, which is suboptimal. If, on the other hand, the candidate thresholds can transmit feasibly until deadline $T$, an empty battery event does not come up until $T$, and an empty battery at $T$ is not possible with an ever decreasing set of thresholds. With these cases ruled out, the decision for an optimal candidate pair can be summarized as follows:

*Lemma 5:* The lowest threshold yielding an empty battery at some time $t_1 < T$ is optimal only if the same threshold, when applied past $t_1$, yields a full battery at some $t_2$ such that $T \ge t_2 > t_1$, or does not yield a battery event until the end of transmission. Conversely, the highest threshold yielding a full battery at some time $\bar{t}_1 < T$ is optimal only if the same threshold, when applied past $\bar{t}_1$, yields an empty battery at some $\bar{t}_2$ such that $T \ge \bar{t}_2 > \bar{t}_1$.

With the restrictions in Lemma 5, the optimal threshold function $p_u(t)$ and the corresponding $p_s(t)$ can be determined by a search algorithm. First, the smallest and largest candidates depleting or filling the battery at some time $t_1$ and $\bar{t}_1$ are found. Out of the two candidates, the one satisfying the relevant condition in Lemma 5 is chosen to as the optimal thresholds. The procedure is then repeated for the next set of thresholds until a feasible set of thresholds depleting the battery at $t = T$ is found, at which point the algorithm terminates.

It is necessary for completeness to point out that for all possible realizations of the candidate thresholds, there exists only and exactly one candidate that satisfies the criteria in Lemma 5. This ensures that the proposed algorithm yields a unique policy, and that it can always find one.

*Lemma 6:* For any pair of candidate thresholds found as the minimum and maximum battery depleting and filling thresholds, there exists exactly one candidate satisfying the corresponding criteria in Lemma 5.

*Proof:* Let the two distinct candidates for the lower threshold be $p_s^{(0)}$ and $p_s^{(E_{max})}$, yielding an empty battery and a full battery at some $t_1$ and $\bar{t}_1$ respectively. Clearly, $p_s^{(0)} > p_s^{(E_{max})}$, since otherwise one of the thresholds violate energy availability or battery capacity constraint at $min(t_1, \bar{t}_1)$. Assume that both candidates satisfy the conditions of Lemma 5, the threshold $p_s^{(0)}$ fills the battery at some $t_2 < T$ or extends feasibly to $T$, and the threshold $p_s^{(E_{max})}$ depletes the battery at some $\bar{t}_2 < T$ by construction. If $\bar{t}_2 < t_1$, the first candidate is not feasible since it must deplete the battery before $\bar{t}_2$ due to $p_s^{(0)} > p_s^{(E_{max})}$. Else if $t_1 \le \bar{t}_2 < t_2$, the first candidate must deplete the battery again before $t_2$, and thus cannot satisfy the conditions of the Lemma. Lastly if $\bar{t}_2 \ge t_2$, then the second candidate yields a battery overflow at $t_2$ and thus is not a feasible candidate. Thus, both candidates cannot satisfy the

$$\mathcal{L} = \int_0^T r(h(t) - s(t) + u(t))\ dt + \int_0^T \lambda(t) \int_0^t \eta s(\tau) - u(\tau)\ d\tau\ dt - \int_0^T \beta(t) \left( \int_0^t \eta s(\tau) - u(\tau)\ d\tau - E_{max} \right)\ dt + \int_0^T \mu(t)(h(t) - s(t))\ dt + \int_0^T \sigma(t) s(t)\ dt + \int_0^T \nu(t) u(t)\ dt \tag{22}$$

conditions of Lemma 5.

Conversely assume that $p_s^{(0)}$ does not satisfy the conditions, i.e., any feasible candidate $p_s^{(0)}$ depletes the battery again at some $t_2$. Then, starting from $p_s^{(0)}$ and decreasing this threshold, one can always find a value at which the battery is full at some $\bar{t}_1$ and is depleted at some $\bar{t}_2 > t_2$. Choosing this value as $p_s^{(E_{max})}$, the second threshold satisfies the required conditions.

Finally, assume that $p_s^{(E_{max})}$ does not satisfy the conditions, i.e., any feasible candidate $p_s^{(E_{max})}$ cannot deplete the battery at any future instance $\bar{t}_2 > \bar{t}_1$. Then, any threshold larger than $p_s^{(E_{max})}$ cannot yield a full battery at any time $t < T$, and therefore the smallest such threshold depleting the battery at some $t_1$ must extend to $T$, making it a feasible candidate for $p_s^{(0)}$. In short, the two candidates cannot satisfy the conditions of Lemma 5 simultaneously, and the failure of either candidate implies the success of the other candidate, proving that there is exactly one candidate that is optimal. ■

Since the problems are mathematically identical, this solution also extends to the broadcast setting.

*Remark 3:* Similar to Remarks 1 and 2, the policy for the finite battery case derived in this section also coincides with previous results for the ideal battery case studied in [3]. When $\eta = 1$ and $p(t)$ is strictly concave, the thresholds are equal and thus the optimal policy is a constant power policy. The power levels increase and decrease at empty and full storage instances, and are chosen analogously to the criterion in Lemma 5 in [3, Theorem 1].

## VI. Online Transmission Policies

In the previous sections, the optimal policy is found by analyzing the harvesting process $h(t)$ over the entire transmission duration $[0, T]$. Therefore it is necessary to know the realization of the energy harvesting non-causally in order to determine the optimal transmission powers, i.e., the policy is found in an offline manner. This approach provides us a benchmark solution as well as insights for efficient power allocation, besides being applicable in some special cases where the harvested energy is highly predictable or controlled. However, such knowledge may not be available in all energy harvesting applications. In this section, we develop online policies that only requires the distribution and causal harvesting information based on the double-threshold structure of the offline policies of earlier sections.

### A. Optimal Online Policy

Without non-causal information of the harvested energy, the transmitter needs to determine its action based only on the current states as well as the previous realizations of the arrival process. The best such policy can be found using dynamic programming [19]. This approach realizes a recursive definition of the value function, with the desired action being the value maximizer, and solves for the optimal action iteratively.

For an energy harvesting transmitter, the states at time $t$ are the energy stored in the battery, $E_{bat}(t)$, causal harvested energy information $h^t = h(\tau)$, $0 \leq \tau \leq t$, and the time to deadline, $T - t$. Based on these states, the node decides on its action, i.e., transmit power, through the function $\phi(E_{bat}(t), h^t, T - t)$. The value function, i.e., expected throughput of the system starting from the given state, is expressed as

$$V(E_{bat}, h^t, T-t) = \max_{\phi} \mathbb{E}\left[\int_t^T r(\phi(E_{bat}(\tau), h^\tau, T - \tau)) d\tau\right] \tag{26}$$

The optimal action $\phi(.)$ is then chosen as the argument that maximizes above value function.

To solve this problem iteratively, we need to express this value function as a recursive relation. Thus, we approximate the integral in (26) as a Riemann sum with interval length of $\delta$. Since the contribution of the next interval is determined by the immediate action $\phi(E_{bat}(t), h^t, t)$, the value function is given by the Bellman equation in (27).

Taking the expectation over the distribution of the harvesting process, this equation can be solved iteratively. However, it is possible to further decrease the dimension of the problem to make it more tractable. For example, if the arrival process is Markovian or i.i.d., the past states do not provide any additional information about the process. Thus, the state $h^t$ can be replaced by only the current harvesting rate $h(t)$. The time until the end of transmission, $T - t$, is helpful towards the very end of the transmission when the node desires to fully consume its energy. For sufficiently large $T$, or for an infinite deadline, this state can be ignored, significantly reducing computational load. In such cases, a discount factor of $\beta$ is added to stabilize the value function. With these assumptions, the Bellman equation in (27) reduces to (28). Notice that the battery state in the value function on the right hand side of (28) can be found from the states at $t$. The battery state changes linearly with $\phi$, with slope $-\eta$ when $\phi(E, h) < h(t)$ and slope $-1$ when $\phi(E, h) \geq h(t)$. This infliction in the state hints to $\phi(E, h) = h(t)$, i.e., transmitting with harvested power, being an optimal action in some cases.

A sample solution to the dynamic program in (28) is given in Figure 4. The figure shows the optimal action $\phi(E, h)$ in an AWGN channel when the harvested power $h$ within an interval of length $\delta$ is distributed independently and uniformly in $[0, 20]$mW, with a battery capacity of $E_{max} = 100$mJ. It can be observed that the optimal transmit power is equal to the harvested energy for a range of states, marked by I,

$$V(E_{bat}, h^t, T-t) = \max_{\phi} \mathbb{E}\left[ r(\phi(E_{bat}, h^t, T-t))\delta + \int_{t+\delta}^{T} r(\phi(E_{bat}(\tau), h^\tau, T-\tau))d\tau \right]$$
$$= \max_{\phi} r(\phi(E_{bat}, h^t, T-t))\delta + \mathbb{E}\left[ V(E_{bat}(t+\delta), h^{t+\delta}, T-t-\delta) \right] \quad (27)$$
$$V(E_{bat}, h) = \max_{\phi} r(\phi(E_{bat}, h))\delta + \beta\mathbb{E}\left[V(E_{bat}(t+\delta), h(t+\delta))\right]. \quad (28)$$

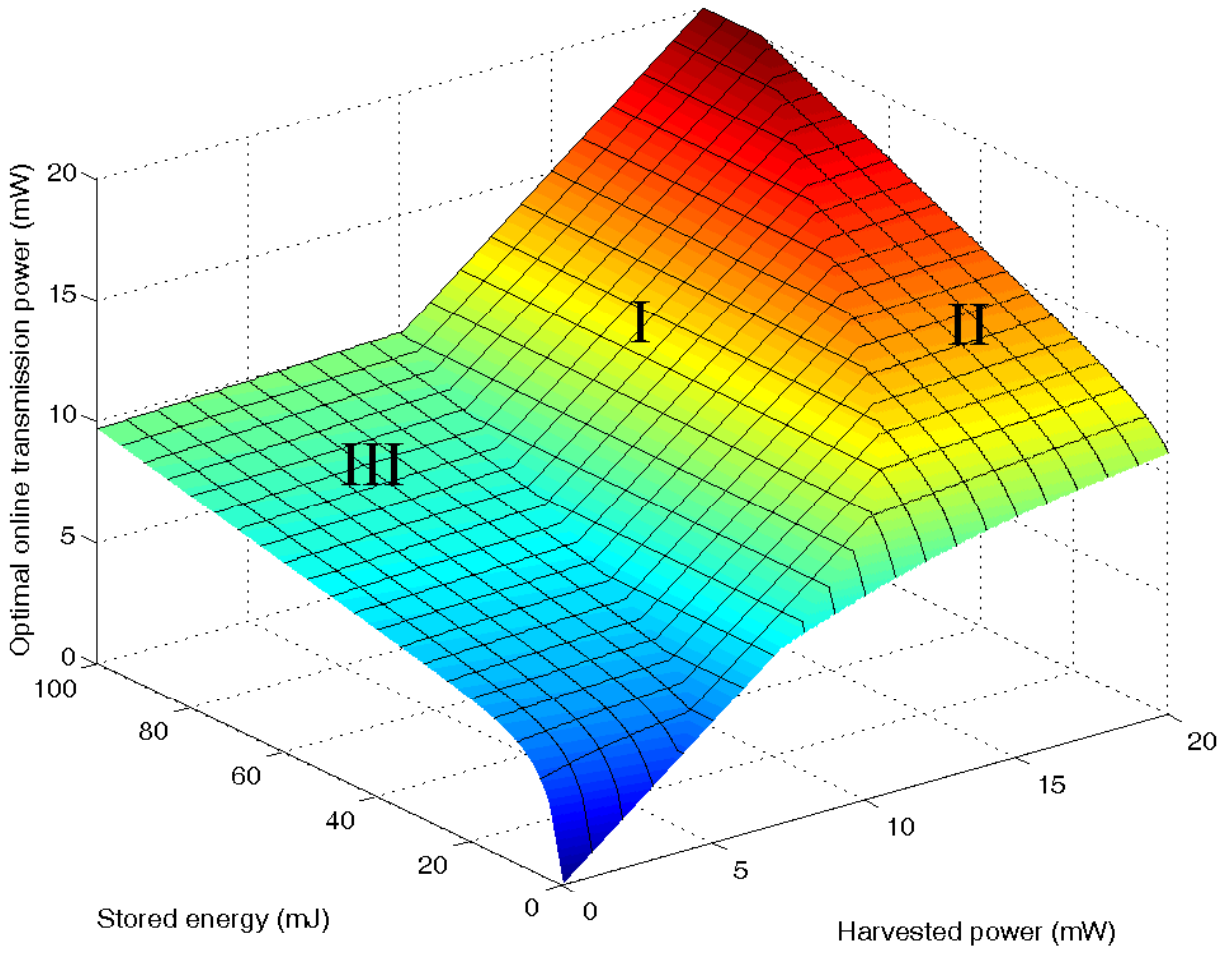

Fig. 4. Optimal online transmission power as a function of node states.

above and below which the transmit powers remain constant (II and III respectively). Moreover, for a fixed stored energy, the two thresholds can be shown to satisfy (15). This aligns perfectly with the optimal offline two-threshold policy, with the thresholds adapting to stored energy rather than non-causally known energy harvests.

### B. Proposed Online Policy

The main insight of the offline policies is that the double threshold structure in Section III and V applies at any time, but the thresholds vary based on the harvesting process. However, in the absence of future harvesting information, it is computationally difficult to estimate the optimum thresholds without solving the dynamic program in Section VI-A. That said, the threshold structure with the relation in (15) can still be utilized with thresholds set independently of future harvesting values. With this in mind, we now propose a simple online policy as follows.

Assuming that the distribution of the harvesting process is known as $f_h(p)$, we propose finding fixed thresholds $p_s(t) = p_s$ and $p_u(t) = p_u$ that simultaneously satisfy

$$\int_{p_s}^{\infty} f_h(p)dp - \int_{0}^{p_u} f_h(p)dp = 0, \qquad \frac{r'(p_s)}{r'(p_u)} = \eta. \quad (29)$$

The first equation in (29) provides long term energy stability by ensuring that the expected energy stored in and drawn from the battery are equal, and thus neither the energy storage is underutilized, nor an excessive amount of energy is stored without utility. Note that as $\eta \to 1$, this reduces to a constant power policy that preserves energy-neutrality, and resembles the best-effort transmission scheme of [20] which is optimal in the information theoretic sense for infinite length transmission. On the other hand at $\eta = 0$, (29) is only satisfied with $p_u = 0$ and $p_s \to \infty$. This means that transmit power is supplied directly by the harvested power, $p(t) = h(t)$, which is optimal since at this efficiency, energy storage is useless. Thus, the proposed online policy achieves the capacity for these two extreme values of $\eta$ in the asymptotical case.

## VII. Numerical Results

To demonstrate the performance of the offline optimal policy and the online policies, in this section we present the numerical results from simulations. Being a more realistic model, we focus on the finite battery case, noting that the resulting insights are similar for the infinite-battery counterpart.

We first focus on a single receiver setting. We consider an energy harvesting transmitter node equipped with a battery of capacity $100mJ$. We assume that the communication channel has Gaussian noise with noise spectral density $N_0 = 10^{-19}W/Hz$ at the receiver, and a bandwidth of $1MHz$. The path loss between the transmitter and receiver is $-100dB$. The transmit duration is taken to be $T = 10000$ seconds. For practical purposes, the continuous model is approximated via sampling at $100$ samples per second. The harvesting process $h(t)$ at each sample point is generated in an i.i.d. fashion, distributed uniformly in $[0, 40]$mW.

Figure 5 shows the average rates achieved with the optimal offline policy of Section V and the online policies of Section VI in comparison with two alternative naive algorithms as a function of storage efficiency $\eta$. The *hasty* algorithm uses up the energy as it is harvested, i.e., $p(t) = h(t)$, and its performance is therefore independent of storage efficiency. This algorithm performs relatively well for small values of $\eta$ as expected, but is surpassed by the others as storage becomes a feasible option. The *constant* algorithm targets a constant transmission level $p_c$ equal to the average harvesting rate. Although optimal for an infinite and efficient battery in the asymptotical case, this algorithm relies significantly on energy storage and therefore fails for smaller values of $\eta$.

As seen in the figure, an efficient battery provides a significant performance advantage in all cases except the hasty policy. The hasty algorithm is optimal at $\eta = 0$, while the constant power policy approaches optimal online with increasing storage efficiency. The proposed online policy performs *at least as good as* both the hasty and constant power algorithms for all values of $\eta$ by mimicking both in the two extreme cases. In this plot, as well as in extensive simulations not presented here, it is observed that the proposed online algorithm performs significantly well in comparison with the online optimal policy, while both remain notably close to the optimal offline upper bound in the absence of non-causal harvesting information.

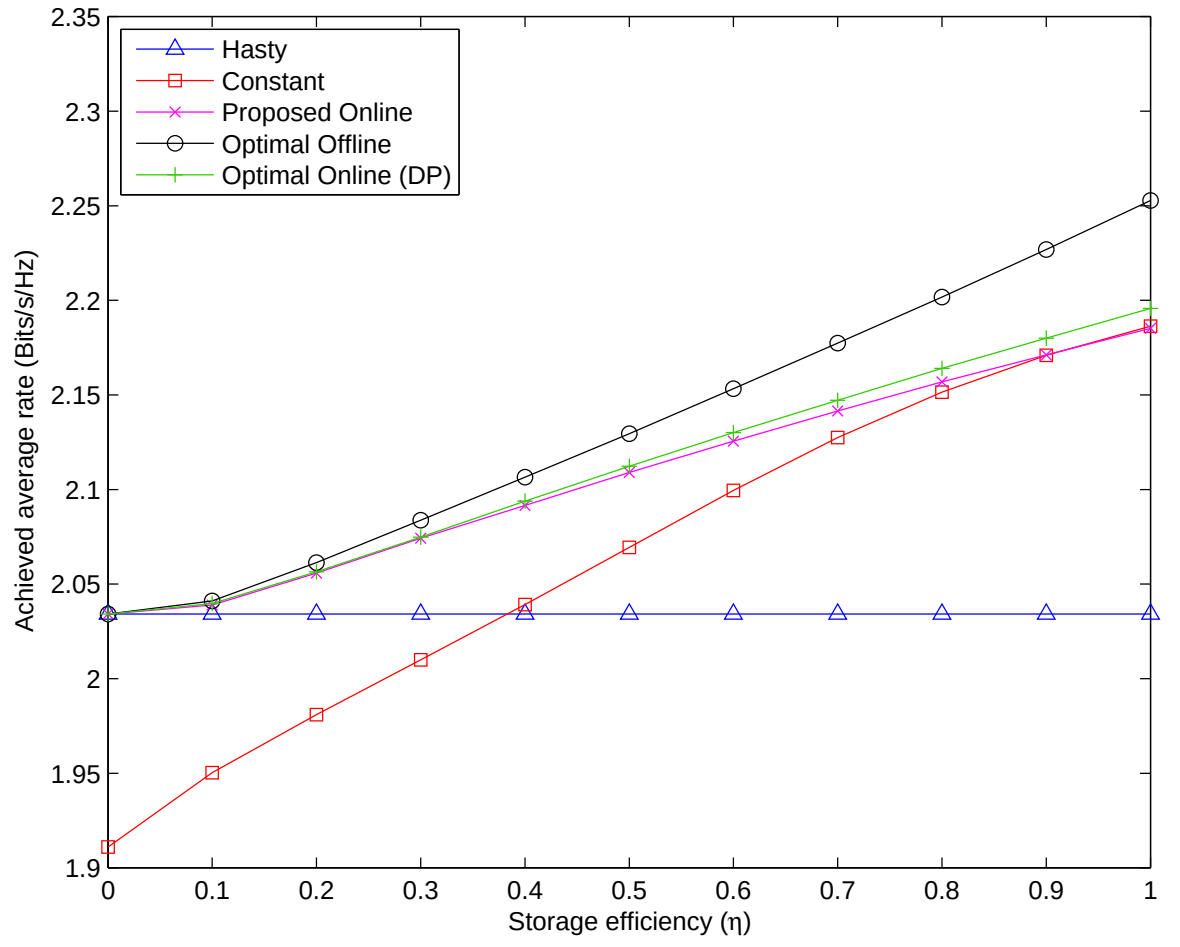

Fig. 5. Average transmission rates versus battery efficiency for the optimal and proposed online algorithm in comparison to naive online algorithms.

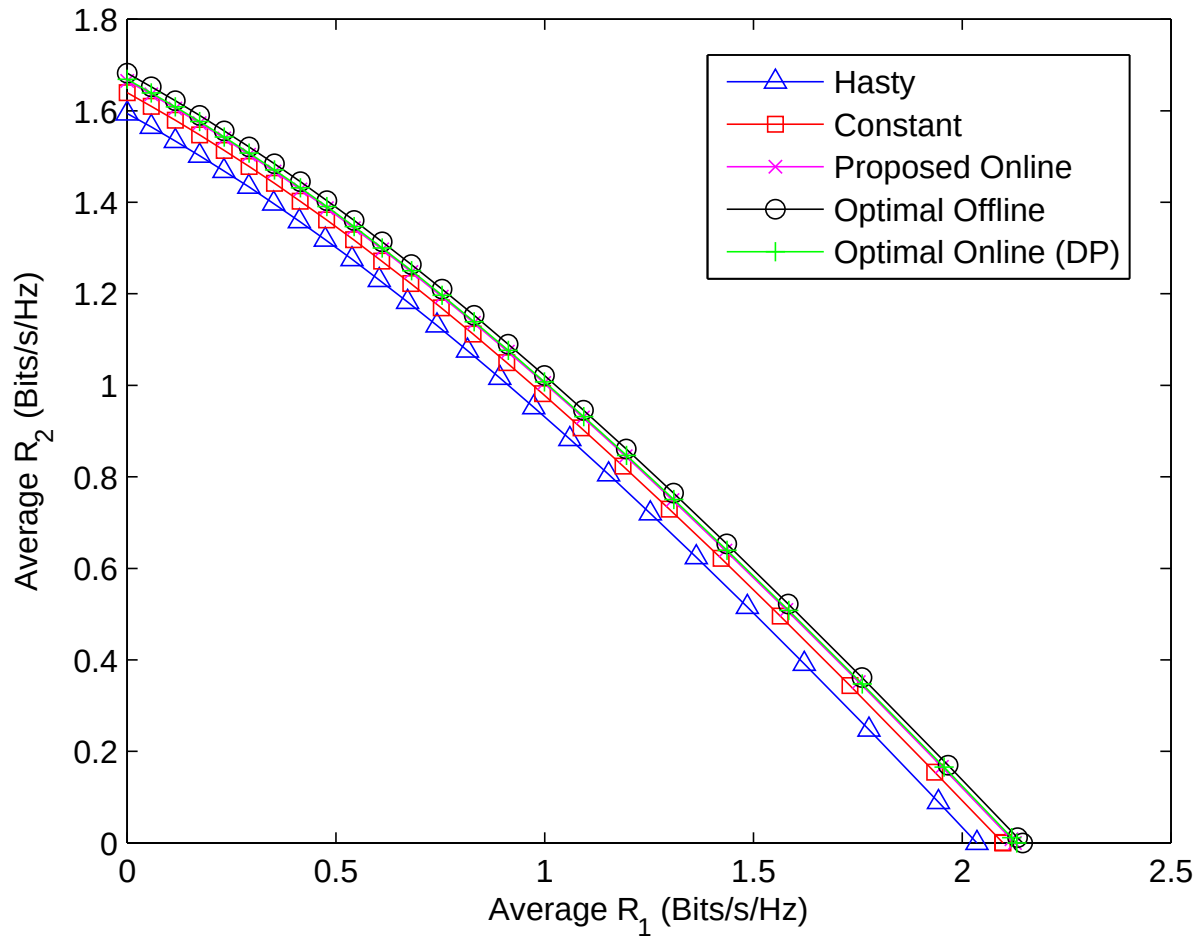

Fig. 6. Average transmission rate regions in an energy harvesting broadcast setting with $\eta = 0.5$.

To observe how these results reflect to the broadcast channel, we perform a simulation with identical parameters and two receivers while fixing the efficiency as $\eta = 0.6$ and plot the achievable average rate region for the two user Gaussian broadcast setting in Figure 6. As described in Section IV, this region is determined by tracing its boundary using a range of values for the ratio $a$. For each value of $a$, the maximum weighted sum is calculated, yielding the average rates $R_1$ and $R_2$ on the boundary with tangent $a$. In comparison to the rates achieved with the naive algorithms, the achievable regions for online and offline policies are shown in Figure 6 when the path loss for the two users are $-100dB$ and $-103dB$ respectively. It is observed that the optimal offline policy allows a larger rate region to be achieved compared to the naive algorithms, while the proposed online algorithm achieves very close to the optimal online boundary and fairly close to the optimal offline boundary.

## VIII. Conclusion

In this paper, the optimal transmit power policy for an energy harvesting transmitter with an inefficient energy storage device was identified. For an infinite battery, it was shown that the optimal policy has a double-threshold structure, where the thresholds are related and are a function of the harvesting process and storage efficiency. The thresholds were shown to be non-decreasing with specific properties that allow them to be found using a simple search algorithm. Using the single user policy, in the broadcast setting the weighted sum rate maximizing policy was shown to have an identical structure. The results were then extended to the case with a finite storage capacity. It was observed that while differing significantly when battery is inefficient, the optimal transmission policies proposed in this paper converges to previous results as efficiency goes to 1. Additionally, the optimal online policy was found using dynamic programming, and was shown to have the two-threshold structure with the thresholds adapting to battery state. Based on the insights from these results, a fixed-threshold online policy was proposed and shown to perform notably well in a single user setting with finite battery capacity compared to other naive power allocation algorithms, while closely tracking the optimal online policy.

An interesting insight of this study is that when battery inefficiency is considered, the optimal power policy is no longer piecewise constant as was the case in previous work with ideal batteries. In fact, in between the two thresholds, the optimal transmitter power turns out to be equal to the harvested power, dictating using up the harvested energy without storing. A relevant future direction is thus developing efficient and practical coding schemes for a transmission with a varying power constraint unknown to the receiver. Another topic of interest is the extension to multiple energy harvesting transmitter scenarios.